\documentclass[b5paper]{appolb}

\usepackage[utf8]{inputenc}
\usepackage[T1]{fontenc}

\usepackage{amsmath}

\usepackage{amsfonts}
\usepackage{mathrsfs}

\usepackage{amssymb}
\usepackage{units}

\usepackage{graphicx}

\usepackage{placeins}

\usepackage{verbatim}

\usepackage{accents}

\usepackage{multirow}

\usepackage{color}

\newcommand{\mrm}{\mathrm}

\makeatletter

\let\old@dmathbeg\[
\let\old@dmathend\]

\newcommand{\rovnec}[1]{\old@dmathbeg#1\old@dmathend}
\newcommand{\rovcis}[2]{\begin{equation}#1\label{#2}\end{equation}}
\newcommand{\drovcis}[2]{\begin{equation}\begin{split}#1\end{split}\label{#2}\end{equation}}
\newcommand{\drovnec}[1]{\begin{equation*}\begin{split}#1\end{split}\end{equation*}} 
\newcommand{\provcis}[1]{\begin{align}#1\end{align}}
\newcommand{\provnec}[1]{\begin{align*}#1\end{align*}}

\newcommand{\rov}{\@ifstar\rovnec\rovcis}
\newcommand{\drov}{\@ifstar\drovnec\drovcis}
\newcommand{\prov}{\@ifstar\provnec\provcis}

\newcommand{\vast}{\bBigg@{4}}
\newcommand{\Vast}{\bBigg@{5}}

\makeatother

\DeclareMathOperator{\sgn}{sgn}

\DeclareMathOperator{\diffbold}{\mathbf{d}}
\newcommand{\bd}{\diffbold\!}

\newcommand{\mca}{\mathcal}

\newcommand{\mbs}{\boldsymbol}

\DeclareFontEncoding{LGR}{}{}

\DeclareMathAlphabet{\msi}{OT1}{cmss}{m}{it}

\DeclareMathAlphabet{\mgr}{LGR}{cmr}{m}{n}

\newcommand{\rpi}{\mgr{p}}

\renewcommand{\[}{\left[}
\renewcommand{\]}{\right]}

\newcommand{\f}{\!\left}
\newcommand{\ri}{\right}

\newcommand{\zrov}{{}\\{}}

\newcommand{\lbl}{\label}
\newcommand{\rvt}{\ .}
\newcommand{\rvc}{\ ,}
\newcommand{\rvs}{\ ;}
\newcommand{\qt}[1]{``#1''}
\newcommand{\gm}[1]{«#1»}

\renewcommand{\(}{\left(}
\renewcommand{\)}{\right)}

\begin{document}
\title{NOTES ON EXTRACTION OF ENERGY FROM\\ AN EXTREMAL KERR--NEWMAN BLACK HOLE\\ VIA CHARGED PARTICLE COLLISIONS\thanks{Presented at the 7\textsuperscript{th} Conference of the Polish Society on Relativity, Łódź, Poland, 20--23 September 2021.}%
}
\author{Filip Hejda
\address{CEICO, Institute of Physics of the Czech Academy of Sciences\\
Na Slovance 1999/2, 182 21 Prague 8, Czech Republic}
\\[3mm]
José P. S. Lemos
\address{CENTRA, Departamento de Física, Instituto Superior Técnico
\\Universidade de Lisboa, Avenida Rovisco Pais 1, 1049-001 Lisboa, Portugal}
\\[3mm]
{Oleg B. Zaslavskii 
\address{Department of Physics and Technology\\
Kharkov V.N. Karazin National University\\
4 Svoboda Square, Kharkov 61022, Ukraine\\
and\\
Institute of Mathematics and Mechanics,
Kazan Federal University\\
18 Kremlyovskaya Street,
Kazan 420008, Russia}
}
}
\headauthor{F. Hejda, J. P. S. Lemos, O. B. Zaslavskii}
\headtitle{Notes on Extraction of Energy from an Extremal Kerr--Newman \ldots}
\maketitle
\begin{abstract}
The so-called BSW effect is an idealised scenario for high-energy test particle collisions in the vicinity of black holes; if the black hole is extremal and one of the particles fine-tuned, the centre-of-mass collision energy can be arbitrarily high. It has been recently shown that the energy of escaping particles produced in this process can also be arbitrarily high in the given approximation, as long as both the black hole and the escaping particles are charged, regardless of how small the black-hole charge might be. We revisit these results and show that they are also compatible with properties of microscopic particles for the case of motion in the equatorial plane of an extremal Kerr--Newman black hole.
\end{abstract}

\section{Introduction}

Bañados, Silk, and West (BSW) \cite{BSW} described a possibility to attain arbitrarily high centre-of-mass collision energies in collisions involving fine-tuned test particles orbiting an extremal Kerr black hole. This BSW effect turned out to be ubiquitous for extremal rotating black holes \cite{Zasl10}, yet it was shown that the energies of particles that can be produced in such a process are subjected to unconditional upper bounds \cite{multi}. On the other hand, no such bounds were found \cite{Zasl12c} for the variant of the BSW effect that can occur for charged particle collisions in the vicinity of an extremal Reissner--Nordström black hole \cite{Zasl11a}. Recently, a more general variant of the BSW effect was considered for charged particle collisions near a general extremal rotating electrovacuum black hole \cite{a2}, unifying the previously known cases. In this setup, it was shown that the unconditional upper bounds on the energy extracted from the black hole by the escaping particles are absent whenever both the black hole and the escaping particles are charged \cite{a4}.

\section{General considerations}

Let us first consider a general axially symmetric, stationary electrovacuum black-hole spacetime with metric
\rov{\mbs g=-N^2\bd t^2+g_{\varphi\varphi}\(\bd\varphi-\omega\bd t\)^2+g_{rr}\bd r^2+g_{\vartheta\vartheta}\bd\vartheta^2\rvc}{axst}
and with the following potential of the electromagnetic field:
\rov{\mbs A=A_t\bd t+A_\varphi\bd\varphi=-\phi\bd t+A_\varphi\(\bd\varphi-\omega\bd t\)\rvt}{aaxst}
The horizon of the black hole corresponds to $N=0$. We further assume that $g_{\varphi\varphi}>0$, that the product $N\sqrt{g_{rr}}$ is finite and nonvanishing for ${N\to0}$, and also that the spacetime has a reflection symmetry with respect to the equatorial \qt{plane} $\vartheta=\nicefrac{\rpi}{2}$. Then, we can consider a motion of charged test particles confined to this hypersurface, with the following equations of motion:
\prov{p^t&=\frac{\mca X}{N^2}\rvc&p^\varphi&=\frac{\omega\mca X}{N^2}+\frac{L-qA_\varphi}{g_{\varphi\varphi}}\rvc&p^r={\frac{\sigma\mca Z}{N\sqrt{g_{rr}}}}\rvt\lbl{eomp}}
Here, the auxiliary functions $\mca X$ and $\mca Z$ are defined by
\prov{\mca X&=E-\omega L-q\phi\rvc&\mca Z=\sqrt{\mca X^2-N^2\[m^2+\frac{\(L-qA_\varphi\)^2}{g_{\varphi\varphi}}\]}\rvs\lbl{XZ}}
\newpage{}
\noindent and $E$ is the particle's energy, $L$ its angular momentum, $q$ its charge, and $m$ its mass. Parameter $\sigma=\pm1$ controls the direction of the radial motion. In order to preserve causality, we need to impose the condition of motion forward in time, \emph{i.e.}, $\mca X>0$.

For an extremal black hole with its horizon located at $r=r_\mrm{H}$, we can write $N^2=\(r-r_\mrm{H}\)^2\mca N^2$. We shall denote values of various quantities at $r=r_\mrm{H}$ with a subscript or superscript $\mrm{H}$. Let us also define the first-order expansion coefficients, $\hat\omega$ and $\hat\phi$, of $\omega$ and $\phi$ by
\prov{\omega&=\omega_\mrm{H}+\hat\omega\(r-r_\mrm{H}\)+\dots\rvc&\phi&=\phi_\mrm{H}+\hat\phi\(r-r_\mrm{H}\)+\dots}

We can observe that only particles with $\mca X_\mrm{H}>0$, called usual, can fall into the black hole. Particles with $\mca X_\mrm{H}=0$ are called critical, as they are on the edge of being able to fall inside. For critical particles, we can approximate functions $\mca X$ and $\mca Z$ close to $r_\mrm{H}$ as follows:
\prov{\mca X&\approx\chi\(r-r_\mrm{H}\)+\dots\rvc&\mca Z&\approx\zeta_\mrm{cr}\(r-r_\mrm{H}\)+\dots}
We can also consider so-called nearly critical particles, which behave approximately as critical at a given radius $r_\mrm{C}$. For such particles we define a formal expansion of $\mca X_\mrm{H}$ in powers of $\(r_\mrm{C}-r_\mrm{H}\)$, 
\rov{\mca X_\mrm{H}\approx-C\(r_\mrm{C}-r_\mrm{H}\)-D\(r_\mrm{C}-r_\mrm{H}\)^2+\dots}{XZcr}
Then the functions $\mca X$ and $\mca Z$ have the following expansions at $r_\mrm{C}$:
\prov{\mca X&\approx\(\chi-C\)\(r_\mrm{C}-r_\mrm{H}\)+\dots\rvc&\mca Z&\approx\zeta_\mrm{nc}\(r_\mrm{C}-r_\mrm{H}\)+\dots}
Here, $\chi$ and $\zeta_\mrm{nc}$ can be expressed as 
\prov{\chi&=-\frac{\hat\omega E}{\omega_\mrm{H}}-q\frac{\omega_\mrm{H}\hat\phi-\hat\omega\phi_\mrm{H}}{\omega_\mrm{H}}\rvc&\zeta_\mrm{nc}=\sqrt{\(\chi-C\)^2-\mca N^2_\mrm{H}\[m^2+\frac{\(E+qA_t^\mrm{H}\)^2}{g_{\varphi\varphi}^\mrm{H}\omega_\mrm{H}^2}\]}\rvc}
whereas $\zeta_\mrm{cr}$ is obtained by putting $C=0$ in $\zeta_\mrm{nc}$.

Let us now consider a collision of two incoming particles, a critical particle $1$ and a usual particle $2$, at a point $r_\mrm{C}$ close to $r_\mrm{H}$. It can be shown that the centre-of-mass collision energy $E_\mrm{CM}$ in such a process goes like $E_\mrm{CM}^2\sim\(r_\mrm{C}-r_\mrm{H}\)^{-1}$ (see \cite{a2, a4}). Let us assume that two new particles, $3$ and $4$, are produced in the event. Now, we shall determine whether the arbitrarily high $E_\mrm{CM}$ can lead to one of their energies being high, too. Let us first note that $N^2p^t\mp N\sqrt{g_{rr}}p^r=\mca X\mp\sigma \mca Z$. For usual particles, it holds $\mca X-\mca Z\sim \(r-r_\mrm{H}\)^2$ and $\mca X+\mca Z\approx2\mca X_\mrm{H}$, whereas for (nearly) critical particles, we have $\mca X\mp\mca Z\sim \(r_\mrm{C}-r_\mrm{H}\)$. Therefore, if we calculate the sum of the momentum conservation laws for $p^t$ and $p^r$ with appropriate coefficients, 
\rov{N^2\(p^t_{1}+p^t_{2}\)\mp N\sqrt{g_{rr}}\(p^r_{1}+p^r_{2}\)=N^2\(p^t_{3}+p^t_{4}\)\mp N\sqrt{g_{rr}}\(p^r_{3}+p^r_{4}\)\rvc}{trsum}
incoming usual particles, outgoing usual particles, and (nearly) critical particles will each have a different leading-order contribution to its expansion in powers of $\(r_\mrm{C}-r_\mrm{H}\)$. We can then infer from \eqref{trsum} with its upper sign that one of the produced particles, say $4$, has to be usual and incoming, and hence bound to fall into the black hole. The other produced particle needs to be nearly critical, as required by \eqref{trsum} with its lower sign, which turns into 
\rov{\chi_1^{\vphantom{\mrm{cr}}}-\zeta_1^\mrm{cr}=\chi_3^{\vphantom{\mrm{nc}}}-C_3^{\vphantom{\mrm{nc}}}+\sigma_3^{\vphantom{\mrm{nc}}}\zeta_3^\mrm{nc}\rvt}{consfin}
One can define a new parameter $\msi A_1$ by $\mca N_\mrm{H}\msi A_1\equiv\chi_1^{\vphantom{\mrm{cr}}}-\zeta_1^\mrm{cr}$ and then solve \eqref{consfin} for $C_3$ and $\sigma_3$. The solution for $C_3$ can be factorised 
\rov{C_3=-\frac{{\mca N}_\mrm{H}}{2g_{\varphi\varphi}^\mrm{H}\omega_\mrm{H}^2\msi A_1}\(E_3-\msi R_+\)\(E_3-\msi R_-\)}{}
in terms of $\msi R_\pm\f(q_3\)$. If we choose $\msi R_+>\msi R_-$, particle $3$ will have $C_3>0$ for $\msi R_+>E_3>\msi R_-$. Since particle $3$ with $C_3>0$ cannot fall into the black hole, it is locally guaranteed to escape. It can be shown that $\msi R_+\f(q_3\)$ is \emph{generically} the highest energy that a particle with a given value of $q_3$ can extract from the black hole (see \cite{a4}).

\section{Selected results for Kerr--Newman solution}

The well-known Kerr--Newman solution describes a charged, rotating black hole with mass $M$, angular momentum $aM$, and charge $Q$. The extremal case is defined by the constraint $M^2=Q^2+a^2$, which leads to $r_\mrm{H}=\sqrt{Q^2+a^2}$. The condition $\mca X_\mrm{H}=0$ for critical particles implies the following relation among $E$, $L$, and $q$:
\rov{E\(Q^2+2a^2\)-aL-qQ\sqrt{Q^2+a^2}=0\rvt}{critkn}
Critical particles can participate in near-horizon collisions only when their parameters have values for which the quantity $\zeta_\mrm{cr}$ is real. Such values form an admissible region in the parameter space of critical particles (see \cite{a2,a4}). Its border is given by the condition $\zeta_\mrm{cr}=0$, which can be expressed as
\rov{q_\pm\f(E\)=\frac{\sqrt{Q^2+a^2}}{Q^3}\[E\(Q^2-a^2\)\pm\left|a\ri|\sqrt{E^2\(Q^2+a^2\)-m^2Q^2}\]\rvt}{admhypqE}
Plugging in $E=m$, we obtain two values
\prov{q_+^\mrm{mb}&=m\frac{\sqrt{Q^2+a^2}}{Q}\rvc&q_-^\mrm{mb}&=m\frac{\sqrt{Q^2+a^2}}{Q^3}\(Q^2-2a^2\)\rvt\lbl{eunit}}
Together with the corresponding value of $L$ calculated from \eqref{critkn}, $q_+^\mrm{mb}$ implies $\mca Z\equiv0$. Hence, the motion is allowed only for values in the vicinity of $q_+^\mrm{mb}$.

The conditional bound $\msi R_+\f(q_3\)$ on extracted energy is given by
\drov{\msi R_+={}&\frac{q_3Q}{\sqrt{Q^2+a^2}}+\frac{1}{Q^2+a^2}\bigg[2a^2\msi A_1+\zrov&+\left|a\ri|\sqrt{\(3a^2-Q^2\)\msi A_1^2+2q_3Q\sqrt{Q^2+a^2}\msi A_1-\(Q^2+a^2\)m_3^2}\bigg]\rvt}{Rpkn}
The parameter $\msi A_1$ takes the form of
\rov{\msi A_1=2E_1-q_1\tilde Q-\sqrt{\(2E_1-q_1\tilde Q\)^2-\[m_1^2+\frac{Q^2+a^2}{a^2}\(E_1-q_1\tilde Q\)^2\]}\rvc}{A1kn}
where $\tilde Q$ is the specific charge of the black hole defined by ${Q=\tilde Q\sqrt{Q^2+a^2}}$.

\section{Conclusions on microscopic particles}

In the case of collisions of particles moving along the axis of symmetry of an extremal Kerr--Newman black hole, it was shown \cite{a3} that even in the absence of unconditional upper bounds on the extracted energy, the process might still be unviable. 
In particular, it was found that for ${\big|\tilde Q\big|\ll1}$, only highly relativistic critical particles were able to approach $r_\mrm{H}$. In the present setup, however, such a problem clearly does not occur, 
since the admissible region contains points with $E<m$ whenever $a\neq0$, as readily seen from \eqref{admhypqE}.
Other caveats studied in \cite{a3} stemmed from the fact that all charged microscopic or elementary particles known in nature have ${\left|q\ri|\gg m}$, and hence the condition \eqref{critkn} generically also implies the initial critical particle $1$ to be highly relativistic. In the equatorial case, this issue can be easily circumvented by considering an uncharged particle $1$, which is very restrictive, however. Thus, we should examine whether $\left|q\ri|\gg m$ can be consistent with a nonrelativistic particle $1$.

We see that both $q_+^\mrm{mb}$ and $q_-^\mrm{mb}$ of \eqref{eunit} 
admit $\left|q\ri|\gg m$, but only for $\big|\tilde Q\big|\ll1$. It is also clear that for $\left|q\ri|\gg m$, 
$q_+^\mrm{mb}$ implies $\sgn q=\sgn Q$, whereas $q_-^\mrm{mb}$ implies $\sgn q=-\sgn Q$. 
We can expand \eqref{admhypqE} for $\big|\tilde Q\big|\ll1$ and use $E\leqslant m$ to obtain two bounds 
corresponding to $q_\pm^\mrm{mb}$; $q\tilde Q<m$ and $-q\tilde Q^3\leqslant2m$. Since charges of microscopic particles are invariable constants of nature, these expressions can be understood as upper bounds on $\big|\tilde Q\big|$ of a black hole which permits processes involving critical microscopic particles with $E\leqslant m$.

Starting with the vicinity of $q_+^\mrm{mb}$, let us now expand $\msi A_1$ of \eqref{A1kn} for $E\approx m$ and ${\left|\tilde q_1\ri|\sim\big|\tilde Q\big|^{-1}\gg1}$ (using $q_1=\tilde q_1m_1$) to obtain
\rov{\msi A_1\approx 2m_1-q_1\tilde Q-\sqrt{2m_1\(m_1-q_1\tilde Q\)}\rvt}{}
We can infer $\msi A_1<m_1$ by utilising $q_1\tilde Q<m_1$. Expanding $\msi R_+$ \eqref{Rpkn} for $\big|\tilde Q\big|\ll1$ and using $\msi A_1<m_1$ therein, we get
\rov{\msi R_+\approx q_3\tilde Q+2m_1+\sqrt{3m_1^2+2q_3\tilde Qm_1-m_3^2}\rvt}{}
If we take $q_1\tilde Q<m_1$ as an upper bound for $\big|\tilde Q\big|$, this changes to
\rov{\msi R_+\approx m_1\(\frac{q_3}{q_1}+2\)+\sqrt{m_1^2\(2\frac{q_3}{q_1}+3\)-m_3^2}\rvt}{}
We see that we can make $\msi R_+$ large only by requiring $q_3\gg q_1$. However, for microscopic particles, it is natural to assume the contrary, $q_1=q_3$, and hence no dependence of $\msi R_+$ on the charges. If we also put $m_3=m_1=m$ for simplicity, we obtain $\msi R_+\approx5m$. 

Turning to $q_-^\mrm{mb}$, let us expand $\msi A_1$ \eqref{A1kn} for $\left|\tilde q_1\ri|\gg\big|\tilde Q\big|^{-1}\gg1$,
\rov{\msi A_1\approx-q_1\tilde Q\rvc}{}
and plug it into $\msi R_+$ \eqref{Rpkn} expanded for $\big|\tilde Q\big|\ll1$ to obtain
\rov{\msi R_+\approx \tilde Q\[q_3-q_1\(2+\sqrt{3-2\frac{q_3}{q_1}}\)\]\rvt}{}
Using $-q_1\tilde Q^3\leqslant2m_1$ as an upper bound for $\big|\tilde Q\big|$ and assuming $\left|q_3\ri|=\left|q_1\ri|=q$ and $q_3=-q_1$, this turns into $\msi R_+\approx\sqrt[3]{2q^2m_1}(3+\sqrt{5})$. Unlike in the previous case, the bound did not lose the dependence on $q$ and it is consistent with $E_3\gg m_3$, \emph{i.e.}, with significant extraction of energy from the black hole.
\bigskip\bigskip

The work of F. H. is supported by the Czech Science Foundation GAČR, Project No. 20-16531Y. J. P. S. L. is grateful for the support from Fundação para a Ciência e a Tecnologia (FCT) provided through Project No.~UIDB/
00099/2020. O. B. Z. thanks the Kazan Federal University for a state grant for scientific activities.

\end{document}